\documentclass[prb,eqsecnum,showpacs,preprint]{revtex4}
\usepackage{epsfig,amssymb,amsmath}
\font\bba=msbm10 scaled 1080
\font\bbb=msbm8 %scaled 1080
\font\bbc=msbm6 %scaled 1080
\newfam\bbfam
\textfont\bbfam=\bba
\scriptfont\bbfam=\bbb
\scriptscriptfont\bbfam=\bbc

\begin{document}
\title{Monte Carlo simulations of the 
critical properties  of the restricted primitive
model}
\author{Jean-Michel Caillol}
\affiliation{Laboratoire de Physique Th\'eorique \\
UMR 8267, B\^at. 210 \\
Universit\'e de Paris-Sud \\
91405 Orsay Cedex, France}
\email{Jean-Michel.Caillol@th.u-psud.fr}
\date{\today}
\begin{abstract}
Recent Monte Carlo simulations of the critical point of the restricted
primitive  model for ionic solutions are reported. Only the continuum version of
the model is considered. A finite size scaling analysis based in the
Bruce-Wilding procedure gives critical exponents in agreement with those of the
three-dimensional Ising universality class. An anomaly in the scaling of the
specific heat with system size is pointed out.
\end{abstract}
\maketitle

\section{Introduction}

 The primitive model (PM) for electrolytes,
 molten salts, collo\"{i}ds, etc is a
 mixture of $M$ species of charged hard spheres living either on a lattice or
 within a continuous volume of real space. In this paper we shall 
 focus only on the off-lattice
 version of the model.
 The simplest version of the PM consists of a binary mixture (\textit{i.e.}
  $M=2$) of  positive  and 
 negative charged hard spheres $\pm q$  all with the same diameter $\sigma$.  
 Under this form the model which is thought to be the  prototype of many 
 ionic fluids has been christened the restricted
 primitive model (RPM). A thermodynamic state of the RPM is entirely specified
 by a reduced density $\rho^*=N \sigma^3/V$ ($N$ number of ions, $V$ volume) 
 and a reduced temperature $T^*=k T \sigma/q^2$ ($k$ Boltzmann's constant).
 
 The RPM undergoes a liquid-vapor transition which has been studied extensively
 these last past years by means of Monte Carlo (MC) simulations and
 various theoretical approaches. 
 In particular the behavior of
 the system  at its critical point (CP) has been the subject of a huge amount of
 numerical and theoretical studies. The question
 is obviously of great importance 
 since it is reasonable to assume that real electrolytes
  - or at last a large class of
 them- and the RPM belong to the same
 universality class which dictates a similar critical behavior. 
 
 It is perhaps the right place to note that
 the main feature of ionic solutions is that the pair potential between two ions
 $i$ and  $j$ at a distance
$r_{ij}=\vert{\bf r}_{i}-{\bf r}_{j}\vert$ which reads as

\begin{eqnarray}
\label{pot}
v (r_{ij}) & = & \frac{q_i q_j}{r_{ij}} \mbox{ for } r_{ij} > \sigma
\nonumber \\
             & = & +\infty    \mbox{ for }  r_{ij} < \sigma
\end{eqnarray} 
is a long range interaction. This fact would suggest  classical (mean field)
behavior,
whereas the well-known screening of the interactions
pleads in favor of  an Ising-like criticality typical
of systems with short range interactions.

On the experimental side it seems
well established now that for many real electrolytes apparent mean field 
behavior applies with sharp crossover (much sharper than in nonionic fluids)
to Ising criticality close to the critical temperature \cite{Schroer}.

At the moment there exists no convincing theoretical proof showing that
the RPM belongs to the Ising
universality class \cite{Fisher1,Stell1,Fisher2,Stell2}
and only sophisticated Monte Carlo simulations can support
this claim. 
Most numerical studies of the CP of the 
off-lattice version of the RPM were performed by the Orsay
group and I would like to  review our contributions towards a better
understanding of the critical properties of this model in the lines below.  
%%%%%%%%%%%%%%%%%%%%%%%%%%%%%%%%%%%%%%%%%%%%%%%%%%%%%%%%%%%%%%%%%%%%%%%%%%%%%%
%%%%%%%%%%%%%%%%%%%%%%%%%%%%%%%%%%%%%%%%%%%%%%%%%%%%%%%%%%%%%%%%%%%%%%%%%%%%
\section{A brief historical survey}
Quite generally,  a single component fluid will undergo a liquid vapor transition 
if the pair potential which is assumed to represent the molecular 
interactions is (sufficiently) attractive at large distances. 
From this point of view  the situation is not so
clear in the case of the RPM (cf eq.\ (\ref{pot}))
and the very existence of the transition is not guaranteed. Several studies were
necessary to clarify this point and a brief historical survey is worthwhile.

The first
evidence that the RPM actually undergoes a liquid-vapor transition  can be
tracked back to two papers of Chasovkikh and Vorontsov-Vel' Yaminov (CVVY)
published as soon as in 1976 \cite{Vorontsov1,Vorontsov2}.
These authors performed isobaric
MC simulations and found a transition with a CP located at 
$T^*_c=0.095$, $\rho^*_c=0.24$.
Several years after (in 1991) Valleau studied three isotherms of the RPM with 
his method of the density scaling MC and obtained a different location for the CP,
namely $T^*_c=0.07$, $\rho^*_c=0.07$ \cite{Valleau}.
Subsequently (in 1992) Panagiotopoulos \cite{Pana1}
obtained  still different results,
\textit{i.e.} $T^*_c=0.056$, $\rho^*_c=0.04$, by performing MC simulations in the Gibbs
ensemble (GE), at the moment a powerful new method of simulation
which he had invented a little bit earlier \cite{Pana2}. 
Subsequent GE simulations 
using an  improved biased MC sampling \cite{Orkou} yielded Panagiotopoulos and
Orkoulas to the new estimate
$T^*_c=0.053$, $\rho^*_c=0.025$.
Finally, making use myself of
the Gibbs ensemble combined with the use of  hyperspherical geometries I
obtained rather $T^*_c=0.057$, $\rho^*_c=0.04$ \cite{Caillol,Caillol1}.

Commenting on this striking dispersion of the MC data  
Prof M. Fisher talked once of the "sad
street of numerical simulations". This was in 1999, at the SCCS
conference, St Malo, France and, 
at this point of the story, I must
agree with him retrospectively. However many advances have been done since.
Before giving an account of these new achievements some comments are in order.
\begin{itemize}

\item{(i)} All the MC studies confirm the existence of a liquid vapor transition for
the RPM. It seems to take place at unusually low densities and temperatures.
Caillol and Weis give further support for such a low critical temperature
\cite{CaillolWeis}.
Moreover it turns out that the coexistence curve is 
very dissymmetric \cite{Pana1,Orkou,Caillol}.

\item{(ii)} The MC simulation of ionic systems is a  numerical challenge due to the
long range of Coulomb potential. In order to deal with this, some caution is
needed. 
Thus, in the case of MC simulations  performed 
in a cubic box with
periodic boundary conditions (PBC), one must use Ewald potentials in order to
obtain the correct physics
\cite{Sahlin,Deleeuw,Caillol2}. The point is that
the Ewald potential is  the solution of Poisson equation in a
cubico-periodical geometry \cite{Caillol2} and many properties of ionic fluids
(electro-neutrality, screening, etc) are a consequence of this fact.
In their  MC simulations CVVY and Valleau
considered truncated Coulomb potentials and  very small samples of $N=32$
particles which yields quantitatively wrong results. By contrast the data of
Panagiotopoulos \textit{et al.} \cite{Pana1,Orkou} are more reliable since
Ewald sums have been used. The same remark apply to
my simulations which were performed on a  4D sphere (a hypersphere for short) 
by considering interactions obtained by solving Poisson equation
in this geometry. This alternative
method of simulation is therefore also indisputably correct,
moreover it is much more efficient. 
The rough agreement observed between 
the simulations of
refs. \cite{Pana1,Orkou} and \cite{Caillol} both involving the same number
of ions, \textit{i.e.} $N=512$, is therefore not fortuitous.  

\item{(iii)} None of the above mentioned studies took correctly
into account finite size effects 
which are of an overwhelming importance near a CP. 
These effects affect the behavior of finite systems as soon as  
the correlation length of the critical density fluctuations is of 
the same order of magnitude  than the size of the simulation box. 
In the simulations \cite{Pana1,Orkou,Caillol} some "apparent" critical
temperature $T_c^*$ has been measured which could be very 
different from its infinite volume limit  $T_c^*(\infty)$.        
\end{itemize}

In order to extract from MC simulations the critical behavior of the RPM in the
thermodynamic limit (\textit{i.e.} the critical exponents)
and also the infinite volume limit of $T_c^*$ and $\rho_c^*$ it is necessary to
perform an analysis of the MC data in the framework of the finite size scaling
(fss) theory which is part of the renormalization group (RG) theory
\cite{Privman,Cardy}. In this approach one needs to work in the Grand Canonical
(GC) ensemble rather than in the Gibbs ensemble which is ill adapted for a fss
analysis.
Subsequent MC simulations on the RPM were thus all performed in this ensemble.
Panagiotopoulos and coworkers turned their attention to the lattice version 
of the RPM whereas the Orsay group continued to work on
its off-lattice version.

%%%%%%%%%%%%%%%%%%%%%%%%%%%%%%%%%%%%%%%%%%%%%%%%%%%%%%%%%%%%%%%%%%%%%%%%%%%%%%%%
%%%%%%%%%%%%%%%%%%%%%%%%%%%%%%%%%%%%%%%%%%%%%%%%%%%%%%%%%%%%%%%%%%%%%%%%%%%%%%%%
\section{Finite size scaling analysis of  MC data }
\subsection{Scaling fields and operators}
Starting with the seminal work of Bruce and Wilding (BW)\cite{BW1,BW2,Wilding}
simulation results for the critical behavior of fluids have customarily been
analyzed along the lines of the so-called revised scaling theory of Rehr
and Mermin \cite{Mermin}. In this approach one 
first defines scaling fields and
operators aimed at restoring the particle-hole symmetry and therefore to map the
the fluid onto a magnetic sytem with Ising-like symmetry. 

The two relevant scaling fields $h$ (the strong ordering field) and $\tau$ (the
weak thermal field) are assumed to be linear combinations of deviations from
their critical values of the chemical potential $\mu$ and the inverse 
temperature $\beta =1/T$ (reduced values are assumed henceforward). 
One thus has
 
\begin{eqnarray}
\label{fields}
h     &=&  \mu - \mu_c + r( \beta - \beta_c) \nonumber \\
\tau  &=&  \beta_c - \beta +s ( \mu - \mu_c) \; ,
\end{eqnarray}
where $r$ and $s$ are the field mixing parameters which define the mapping. Of
course relations\ (\ref{fields}) are valid only in the vicinity
of the CP. The
conjugate scaling operators ${\cal M}$ and ${\cal E}$ are then defined as
\begin{eqnarray}
\label{operators}
<{\cal M} >& = &
\frac{1}{V} \frac{\partial}{\partial h} \ln \Xi =
\frac{1}{1-sr} \ (<\rho> - s <u>)  \nonumber \\
 <{\cal E}>  & = &
\frac{1}{V} \frac{\partial}{\partial \tau} \ln \Xi =
\frac{1}{1-sr} \ (<u> - r <\rho>)   \; , 
\end{eqnarray}
where $ \Xi  $ is the  GC  partition function of the RPM,
$ \rho $ the total number density, and $ u $ the internal energy per unit
volume. Brackets $< \ldots >$ denote GC averages.
${\cal M}$ is the order parameter (magnetization)
of the  magnetic system associated with the fluid and  ${\cal E}$ 
its  magnetic energy. ${\cal E}$ should be
invariant under the transformations $({\cal M},h) \rightarrow (- {\cal M},-h)$
for appropriate values of $s$ and $r$. In this framework the coexistence curve
is therefore defined by the eq.~$h=0$.

The revised scaling of Rehr and Mermin implies the analyticity of the
coexistence chemical potential $\mu(T)$ at
$T_c^*$. Although this is the case for some peculiar lattice gas models with "hidden"
symmetries there is no reason
that in general, for fluid systems $\mu(T)$ should lack a singularity as
recognized already by Rehr and Mermin \cite{Mermin} and emphasized more recently
by Fisher and co-workers \cite{Fisher3,Fisher4,Fisher5} .
%%%%%%%%%%%%%%%%%%%%%%%%%%%%%%%%%%%%%%%%%%%%%%%%%%%%%%%%%%%%%%%%%%%%%%%%%%%%%%%
%%%%%%%%%%%%%%%%%%%%%%%%%%%%%%%%%%%%%%%%%%%%%%%%%%%%%%%%%%%%%%%%%%%%%%%%%%%%%%%
\subsection{The scaling hypothesis }
A central role in the subsequent fss analysis is played by the GC joint 
distribution $\mathcal{P}_L(\mathcal{M},\mathcal{E})
\propto \mathcal{P}_L(\rho,u)$ for the scaling operators
${\cal M}$ and ${\cal E}$. Following BW \cite{BW1,BW2,Wilding}
 we will assume that, in the
immediate vicinity of the CP, 
$\mathcal{P}_L(\mathcal{M},\mathcal{E})$ obeys to the following scaling law:
 
\begin{eqnarray}
\label{scaling}
\mathcal{P}_L({\cal M,E})& = & a_{\cal M}^{-1} \, a_{\cal E}^{-1} \,
 L^{d-y_\tau} \, L^{d-y_h}  
  \widetilde{\mathcal{P}}
   ( a_{\cal M}^{-1} L^{d-y_h} \, \delta {\cal M},\, \nonumber \\ 
    &\ldots &
 a_{\cal E}^{-1} L^{d-y_\tau} 
  \, \delta {\cal E}, a_{\cal M} L^{y_h} \, h, 
 a_{\cal E} L^{y_\tau} \, \tau, a_i L^{y_i},\ldots) \; ,
\end{eqnarray}
where $L$ are the linear dimensions of the system (taken as $V^{1/3}$, where $V$
is the volume of the simulation box, either a cube or a hypersphere). I have
denoted by $\delta {\cal M}\equiv {\cal M} - <{\cal M}>_c$ and 
$\delta {\cal E}\equiv {\cal E} - <{\cal E}>_c$ the deviations of the scaling
operators from their value at criticality.
The cornerstone of this scaling hypothesis is that the function 
$\widetilde{\mathcal{P}}$ which enters eq.\ (\ref{scaling}) is
{\it universal} in the sense that it depends only upon the universality class
of the model and of the type of geometry considered.
The constants $a_{\cal M}$, $a_{\cal E}$, and $a_i$
are system dependent constants which are
defined in such a way that $\widetilde{\mathcal{P}}$  has unit variance.
Finally, the renormalization exponents $y_h$, $y_\tau$, and $y_i$ which enter
eq.\ (\ref{scaling}) are defined as
\begin{eqnarray}
y_h   &= &  d-\beta/\nu  \nonumber \\  
y_\tau & = &   1/ \nu \nonumber \\ 
y_i &=& - \theta/\nu
\end{eqnarray}
in terms of the  usual critical exponents: 
\begin{itemize}
\item $\beta$ exponent of the
ordering field, \textit{i.e.}
 $<\delta {\cal M}>\sim \vert \tau \vert^{\beta} $
for  $T^*<T_c^*$ at $h=0$
\item $\nu$ exponent of the correlation length, \textit{i.e.}
$\xi \sim \vert \tau \vert^{-\nu} $
\item $\theta$ Wegner's correction-to-scaling exponent (first
irrelevant exponent).
\end{itemize}
The scaling hypothesis\ (\ref{scaling}) was established on a solid RG basis
for Ising-like systems \cite{Bruce} and received  substantial supports 
from MC studies
\cite{Nicolaides}. We stress once again that the coexistence curve is 
determined in this approach by the
condition $h=0$ and that, at coexistence, the order parameter distribution
$\mathcal{P}_L(\mathcal{M})$ should be an even function of $\mathcal{M}$. In
practice this symmetry requirement can be satisfied by tuning the 
two parameters ($\mu, s$) at a given $\beta$.
We now concentrate our attention on the scaling
behavior of the histogram $\mathcal{P}_L(\mathcal{M})$.

%%%%%%%%%%%%%%%%%%%%%%%%%%%%%%%%%%%%%%%%%%%%%%%%%%%%%%%%%%%%%%%%%%%%%%%%%%%%%%%
%%%%%%%%%%%%%%%%%%%%%%%%%%%%%%%%%%%%%%%%%%%%%%%%%%%%%%%%%%%%%%%%%%%%%%%%%%%%%%%
\subsection{The matching procedure}

Integrating both sides of eq.\ (\ref{scaling})
over ${\cal E}$ one finds that, along the coexistence line  $h=0$ one has
\begin{eqnarray}
\label{PM}
\mathcal{P}_L({\cal M}) = a_{\cal M}^{-1} \, L^{d-y_h}
\widetilde{\mathcal{P}} \, ( a_{\cal M}^{-1} L^{d-y_h}
 \, \delta {\cal M}, 
 a_{\cal E} L^{y_\tau} \, \tau, a_i L^{y_i}) \; ,
\end{eqnarray}
where, in the r.h.s. the dependence of the universal function
 $\widetilde{\mathcal{P}}$ upon $h$ has been discarded for clarity.
 Let us define now 
 $ x = a_{\cal M}^{-1} L^{d-y_h} \delta {\cal M}$, then,
 assuming $ \tau \sim 0$ and $ L \sim \infty$
 a Taylor expansion
 of eq.\ (\ref{PM}) yields  
   
\begin{eqnarray}
\label{scalingbis}
\mathcal{P}_L({\cal M})  & =  &a_{\cal M}^{-1} \, L^{d-y_h}
\Big[ \widetilde{\mathcal{P}}^*(x)+ a_{\cal E} L^{y_\tau} \,
\tau \ \widetilde{\mathcal{P}}_ 1^*(x)  \nonumber \\
 & + & a'^2_{\cal E} L^{2y_\tau} \, 
 \tau^2 \widetilde{\mathcal{P}}_ 2^*(x)+ ... + 
 a_i L^{y_i}\widetilde{\mathcal{P}}_ 3^*(x) \  +    ...  \Big] \; ,
\end{eqnarray} 
where the various $ \widetilde{\mathcal{P}}^*$ entering the r.h.s. 
are universal functions.
 Note that, for  $L=\infty$ the normalized ordering field distribution 
 $\mathcal{P}_L(\mathcal{M})$
 collapses onto an universal function 
 $\widetilde{\mathcal{P}}^*(x)$ at  $\tau=0$.
 For  $L$ finite but large
 $\mathcal{P}_L(\mathcal{M})$
 collapses approximately onto  $\widetilde{\mathcal{P}}^*(x)$ at
 some apparent $ \tau_L \propto L^{-y_\tau + y_i}$.  Since for $h=0$
 one has $\tau \propto \beta - \beta_ c$ then the matching of the histogram
 $\mathcal{P}_L({\cal M})$ onto the universal function 
 $\widetilde{\mathcal{P}}^*(x)$ should occur at some apparent temperature
 $T_c^*(L)$ scaling with system size as
 \begin{eqnarray}
 \label{Tinf}
T_c^*(\infty) \, - \, T_c^*(L) \propto L^{-(\theta+1)/\nu} +  .... \; ,
\end{eqnarray}
where $T_c^*(\infty)$ denotes the infinite volume limit 
of the critical temperature.

%%%%%%%%%%%%%%%%%%%%%%%%%%%%%%%%%%%%%%%%%%%%%%%%%%%%%%%%%%%%%%%%%%%%%%%%%%%%%%%
%%%%%%%%%%%%%%%%%%%%%%%%%%%%%%%%%%%%%%%%%%%%%%%%%%%%%%%%%%%%%%%%%%%%%%%%%%%%%%%
\subsection{Technical details }
To assess the critical behavior and the critical parameters of the system, we
need, in a first step, to locate the coexistence curve $h=0$. At a given
temperature $\beta$ close to $\beta_c$
the ordering distribution function 
$\mathcal{P}_L(\mathcal{M})$ depends solely on the chemical potential $\mu$ and
the mixing parameter $s$. At coexistence, the value of  ($\mu$,$s$) can be
obtained unambiguously by requiring that $\mathcal{P}_L(\mathcal{M})$ is
symmetric in $\mathcal{M}-<\mathcal{M}>$ \cite{Wilding}. Tuning at will $\mu$
and $s$ at given $\beta$ requires to know the joint  histogram 
$\mathcal{P}_L(\mathcal{M},\mathcal{E})
\propto \mathcal{P}_L(\rho,u)$ for a continuous set of values of $\mu$ at a given
$\beta$.
Moreover, since this analysis must be performed at
different $\beta$ one needs in fact to know
$\mathcal{P}_L(\mathcal{M},\mathcal{E})$ for a continuous set of values of 
($\beta$,$\mu$) in the critical region. 
This technical difficulty is circumvented
by using the multiple histogram reweighting proposed by Ferrenberg and Swensen
\cite{Ferren1,Ferren2,Deutsch}. With this method one can obtains 
$\mathcal{P}_L^{\beta,\mu}(\rho,u)$ for a continuous set of values of 
($\beta$,$\mu$) from the knowledge of $R$ histograms 
$\mathcal{P}_L^{\beta_i,\mu_i}(\rho,u), i=1,\ldots, R$ obtained
by performing $R$ distinct MC simulations in the $R$ (neighbor) 
thermodynamic states ($\beta_i,\mu_i$).

Since the precision of the simulations of fluid systems has still not reached 
that obtained in the MC simulations of Ising like systems it is impossible to
construct {\it ex nihilo} the fixed point universal
distribution $\widetilde{\mathcal{P}}^*(x)$. In refs.~\cite{OR1,OR2} 
our attempts to match 
$\mathcal{P}_L(\mathcal{M})$ on $\widetilde{\mathcal{P}}^*(x)$ were realized by
using the estimate of $\widetilde{\mathcal{P}}^*_{is}(x)$ made by Hilfer and
Wilding \cite{Hilfer} for the 3D Ising model. Two new -and better- estimates of 
$\widetilde{\mathcal{P}}^*_{is}(x)$ obtained by Tsypin and Bl\"ote 
\cite{Tsypin} for the 3D
Ising model and the spin-1 Blume-Capel model were considered in 
ref.~\cite{OR3}. The discussion is postponed to next section.

%%%%%%%%%%%%%%%%%%%%%%%%%%%%%%%%%%%%%%%%%%%%%%%%%%%%%%%%%%%%%%%%%%%%%%%%%%%%%%%
%%%%%%%%%%%%%%%%%%%%%%%%%%%%%%%%%%%%%%%%%%%%%%%%%%%%%%%%%%%%%%%%%%%%%%%%%%%%%%

\section{Results}
\subsection{General discussion}

 It turns out that the field mixing parameter $s$ of the RPM is practically 
 independent of the temperature and of the size $L$ of the system. 
 Its magnitude, $s \sim -1.46$ \cite{OR1,OR2,OR3}, is much higher than for
 neutral fluids (typically  $s \sim 0.02$ for square well or Lennard-Jones
 fluids \cite{CaillolLJ}) which explains the large dissymmetry of the 
 liquid-vapor coexistence curve of the RPM.
 
 The collapse of the ordering operator distribution
  $\mathcal{P}_L(\mathcal{M})$
 onto the universal ordering distribution
 $\widetilde{\mathcal{P}}^*_{is}(x)$ given by the Blume-Capel model
 \cite{Tsypin} is depicted in Fig.~1 for four
 different values of the volume ranging from $V/\sigma^3=5000$ to 
 $V/\sigma^3=40000$, \textit{i.e.} up to a linear size $L/ \sigma=34$.
  At volume $V/\sigma^3=5000$ a mismatch is observed at the
 lowest values of $\mathcal{M}$ due to an inadequate sampling of of the low
 density configurations at small volume. The overall good agreement leads us to
 conclude that the universality class of the RPM is that of the 3D Ising model.
 
 The reduced apparent critical temperature $T_c^*(L)$ versus the size $L$ 
 of the system (in reduced units) has been plotted in Fig.~2. 
 Depending on the choice made for the universal ordering distribution
 $\widetilde{\mathcal{P}}^*_{is}(x)$ one obtains two sets of values 
 of $T_c^*(L)$ from which $T_c^*(\infty)$ can be obtained by making use of eq.\
 (\ref{Tinf}). One obtains $T_c^*(\infty)=0.04917 \pm 0.00002$ using
 $\widetilde{\mathcal{P}}^*_{is}(x)$ derived from the Blume-Capel model and
 $T_c^*(\infty)=0.04916 \pm 0.00002$ using
 $\widetilde{\mathcal{P}}^*_{is}(x)$ obtained for the 3D-Ising model. The
 approximate $\widetilde{\mathcal{P}}^*_{is}(x)$ of Hilfer and Wilding yields
 slightly different results. Note that in all cases we have used the Ising
 values $\nu=0.630$ \cite{Nu} and $\theta=0.53$ \cite{Theta} 
 of the critical exponents. 
 
 The previous analysis merely establishes the compatibility of the MC data with
 an Ising-like criticality. One can try to go beyond by considering the scaling
 behavior of the Binder cumulant
 \begin{eqnarray}
 \label{Q}
 Q_B(L) = \frac {<\delta{\cal M}^2>^2_L} {<\delta{\cal M}^4>_L} \; .
\end{eqnarray}
As a consequence of the scaling hypothesis~(\ref{scalingbis}) one can show
that, at coexistence ($h=0$), $Q_B(L)$ should scale with system size as
 
 \begin{eqnarray}
 \label{Binder}
 Q_B(L) & \ = \ & Q_c + q_1 (\beta - \beta_c)L^{1/\nu} +
  q_2 (\beta - \beta_c)^2 L^{2/\nu} \nonumber \\
  &+& q_3 (\beta - \beta_c)^3 L^{3/\nu} ...
   \ + \  b_1 L^{y_i} + ...   
\end{eqnarray}
 where the last term takes into account  contributions from  irrelevant fields
 and $q_1$,  $q_2$,  $q_3$, and $b_1$ are non universal constants. If the
 contribution of irrelevant fields could be neglected then the curves $Q_B(L)$
 would intersect at the fixed point $ Q_c$. As apparent in Fig.~3 this
 is clearly not the case  and corrections to scaling must be taken into account.
 
 Recall that for the 3D-Ising model the fixed point value  is
 $ Q_c=0.623$ \cite{Q} and that the exponent of the correlation length has the
 value $\nu=0.630$ \cite{Nu}.
 We have attempted to fit all our MC data with eq.\ (\ref{Binder}).
 If all the parameters in the RHS of eq.~(\ref{Binder}) are kept free
 such an ambitious fit turns out to be impossible. 
 Various other less satisfactory strategies can be considered however.    
\begin{itemize}
\item
Fixing $\beta_c=1/0.04917$,
$y_i=-\theta/\nu=-0.84$ and leaving free all the other parameters
one finds a fit better than  $ 1 $ per cent
and $Q_c= 0.63 \pm 0.01$,  $ \nu \sim 0.66 \pm 0.03$.
\item Conversely, fixing $  Q_c= 0.623$ and $\theta=0.53$ one
obtains $\beta_c=1/0.04918$ and  $ \nu \sim 0.63 \pm 0.03$.
\end{itemize} 
The variations of   $Q_B(L)$ as a function of $\beta$ for the different volumes
is shown in Fig.~3. Although there is considerable spread in the intersection
points due to correction-to-scaling contributions, the corresponding values of 
$Q_c $ are close to the Ising value $ Q_c=0.623$ and permit to rule out mean
field behavior (\textit{i.e.} $ Q_c=0.457$ \cite{MFQ}). 

Further support for Ising
criticality is provided by the behavior
of $ <\delta {\cal M}^2>$ at $T^*_c(L)$.
According to the scaling hypothesis~(\ref{scalingbis}) it
should scales as $L^{2 \beta / \nu}$ with system size. From the slope of the
curve displayed in Fig.~4 one obtains $\beta / \nu=0.52$ in accord with the 3D
Ising value (0.517) and in clear contrast with the classical value 1.
 
In summary, our fss analysis leads to an estimate of the critical exponents
$\nu$ and $\beta/\nu$ and the Binder cumulant $Q_c $ based on the sole
 knowledge of
the critical temperature and the renormalization exponent $\theta$. Within the
numerical uncertainties these values are compatible with Ising-like criticality.
Our conclusion is that the RPM, as ordinary neutral fluids,
belongs to the universality class of the Ising model.

A complete discussion of our MC data is out of the scope of the present paper
and can be found in ref.~\cite{OR3}. For completeness I give below the values 
obtained for the
critical temperature, chemical potentials and densities 
(the critical pressure is largely unknown):

\begin{itemize}
\item $T^{\star}_c=0.049 17\pm 0.000 02$
\item $\rho^{\star}_c=0.080 \pm 0.005$
\item $\mu^{\star}_c=-13.600\pm0.005$
\end{itemize}
%%%%%%%%%%%%%%%%%%%%%%%%%%%%%%%%%%%%%%%%%%%%%%%%%%%%%%%%%%%%%%%%%%%%%%%%%%%%%%%
%%%%%%%%%%%%%%%%%%%%%%%%%%%%%%%%%%%%%%%%%%%%%%%%%%%%%%%%%%%%%%%%%%%%%%%%%%%%%%%
    
\subsection{The specific heat}
The revised scaling theory of Rehr and Mermin 
which is the framework of our fss analysis is however not the most general
scaling theory which can be proposed for a fluid system lacking the
"particle-hole" symmetry.
Its main weakness, as recognized already by
Rehr and Mermin \cite{Mermin}, Yang and Yang \cite{Yang}, and more recently by
Fisher and co-workers \cite{Fisher3,Fisher4,Fisher5}, 
is that it assumes the analyticity of the chemical
potential at coexistence $\mu(T)$ at the critical point. The more general
scaling assumption should yield singularities for both  $\mu(T)$ and $p(T)$ as
$T^* \to T_c^*$. Let us examine the consequences of these singularities on the
behavior of the specific heat capacity at constant volume
$C_V$. In the two phase region it can be rewritten as \cite{Yang}

\begin{eqnarray}
\label{Cv}
C_V & =&  V T \left . \frac{\partial^2 p}{\partial T^2}\right |_V -
 N T \left . \frac{\partial^2 \mu}{\partial T^2}\right |_V \nonumber \\
 &=& C_p + C_{\mu}  \; ,
\end{eqnarray} 
where $C_p $ (not to be confused with the specific heat capacity at constant
pressure) and $C_{\mu}$ (not to be confused with the specific heat capacity 
at constant chemical potential) denote the two contributions to $C_V$. I stress
that, in eq.~(\ref{Cv}) $p(T)$ and $\mu(T)$ denote the pressure and the chemical potential
at coexistence. The formula can be used for any density $\rho_g(T) < \rho <
\rho_l(T) $ within the two phase region ($\rho_g(T)$ and $\rho_l(T)$ being the
densities of the gas and the liquid at coexistence respectively).
In the
revised scaling theory only $C_p $ diverges as $\vert 
T^* -T_c^* \vert^{-\alpha}$
whereas one expects a divergence of both $C_p $ and $C_{\mu}$ (both as 
$\vert T^* -T_c^* \vert^{-\alpha}$). In Fig.~5 I display the curves
 $C_{\mu}(T)$ and $C_V(T) $ along the locus 
\begin{equation} 
\chi_{NNN}\propto <(N-<N>)^3>=0 \; ,
\end{equation}
for the four volumes considered in our last MC simulations \cite{OR3}. 
Although the peak positions shift correctly as $\propto L^{-1/\nu}$ with system
size, in accord with fss theory \cite{Privman,Cardy}, there is no detectable
scaling of the heights of the peaks which should scale as 
$L^{\alpha/\nu}$ with $L$. These observations corroborate similar results obtained
by Valleau and Torrie\cite{Valleau1,Valleau2}.
In particular $C_{\mu}$ does not show any anomaly which should
challenge the use of eqs.~(\ref{fields}) for the scaling fields.
A possible explanation for the non singular behavior of $C_V(T)$ is that the
amplitude of the singular term in $C_V(T)$ is small in the RPM and that the
specific heat is dominated by its regular part.
Note however that, the peak heights in $C_V(T)/V$  would scale, assuming
Ising value for $\alpha$ only by a factor $2^{\alpha/\nu}\sim 1.12$ when
doubling the linear dimensions of the system. It is possible that such a small
effect is not detectable within the statistical uncertainties of our
calculations. 
 
%%%%%%%%%%%%%%%%%%%%%%%%%%%%%%%%%%%%%%%%%%%%%%%%%%%%%%%%%%%%%%%%%%%%%%%%%%%%%%%
%%%%%%%%%%%%%%%%%%%%%%%%%%%%%%%%%%%%%%%%%%%%%%%%%%%%%%%%%%%%%%%%%%%%%%%%%%%%%%%
\section{Conclusion} 
 In this paper which resumes my talk at the Lviv NATO workshop I have described
 recent attempts to elucidate the nature of the critical behavior of the
 RPM model for ionic fluids, prototype of a system governed by long
 range Coulomb interactions by means of MC simulations.
 After endeavor over more than a decade we have now
 reached a point where we can claim confidently that the RPM belongs to the same
 universality class than the 3D Ising model. The critical values of
 non-universal quantities such as the temperature and the chemical potential 
 were established with a high accuracy whereas the uncertainties 
 on the critical density are more significant, and the critical pressure is
 unknown. 
 
 The behavior of the constant volume specific heat gives no indication of the
 expected $L^{\alpha / \nu}$ scaling of the peak height within the range
  of system sizes
 considered in the most recent simulations. Recent investigations of Camp and
 co-workers \cite{Camp} where differences in the behavior of  $C_V$ in
 the canonical and the GC ensemble are reported have emphasized this 
 problem. At the
 moment it is difficult to explain this unexpected behavior of the specific
 heat.

 I have only discussed the properties of the
 continuous version of the RPM. The phase diagram of the various lattice
 versions of the model is in fact more complex \cite{Luijten,Panalast} and was
 not described here due to a  lack of place.
 I have also excluded from my presentation 
 assymetric versions, either in charge or/and in size, of the continuum or 
 lattice versions of the primitive model. 
 The interested reader should consult recent works of Panagiotopoulos
 \textit{et al.} \cite{assym1,assym2,assym22} 
 and  de Pablo \textit{et al.} \cite{assym3,assym4,assym5}.

%%%%%%%%%%%%%%%%%%%%%%%%%%%%%%%%%%%%%%%%%%%%%%%%%%%%%%%%%%%%%%%%%%%%%%%%%%%%%%%
%%%%%%%%%%%%%%%%%%%%%%%%%%%%%%%%%%%%%%%%%%%%%%%%%%%%%%%%%%%%%%%%%%%%%%%%%%%%%%%
\section*{Acknowledgments} 
I thank the organizers of the NATO workshop "Ionic Soft Matter" held in Lviv,
Ukraine in April 2004, Pr D. Henderson and Pr M. Holovko for having invited me
to give a talk. It is a pleasure to acknowledge many scientific discussions with
Pr. I. M. Mryglod, O. Patsahan, J.-P.
Badiali, P. J. Camp, W. Schr\"oer, and J. Stafiej.

\newpage
\noindent \textbf{Figure Captions}
\begin{itemize} 
\item \textbf{figure 1:}
Collapse of the ordering  distribution
 $\mathcal{P}_L(\mathcal{M})$ onto the universal Ising ordering distribution
 $\widetilde{\mathcal{P}}^*_{is}(x)$ for  
 $V/\sigma^3=5000$, $T_c^*(L)=0.0049 34$, $s=-1.45$;
 $V/\sigma^3=10000$, $T_c^*(L)=0.0049 26$, $s=-1.465$;
 $V/\sigma^3=20000$, $T_c^*(L)=0.0049 21$, $s=-1.47$; and
 $V/\sigma^3=40000$, $T_c^*(L)=0.0049 22$, $s=-1.43$. 
 $\widetilde{\mathcal{P}}^*_{is}(x)$ (solid circles) if the MC result of Tsypin
 and Bl\"ote (Ref.~\cite{Tsypin}) obtained for the Blume-Capel model.
 The scaling variable is $x=a_{\cal M}^{-1} L^{\beta/\nu}
 \delta {\cal M}$ where 
 $a_{\cal M}$ is chosen in such a way that  $\mathcal{P}_L(x)$ 
 has unit variance.
 
\item \textbf{figure 2:} The apparent critical temperature $T_c^*(L)$ as a
function of $L^{-(\theta+1)/\nu}$ with $\theta=0.53$, $\nu=0.630$ obtained by
matching the universal ordering distribution calculated for the Blume-Capel
model (top) and the Ising model (bottom). Extrapolating by linear least square
fit to the infinite volume limit yields $T_c^*(\infty)=0.049 17 \pm 0.000 02$
(top) and $T_c^*(\infty)=0.049 16 \pm 0.000 02$  (bottom).

\item \textbf{figure 3:} Variation of  $Q_B(L)$ as a function of the inverse
temperature $\beta$ for the different volumes considered in ref.~\cite{OR3}.
From top to bottom $V/\sigma^3=40000,20000,10000$, and $5000$ respectively. The
symbols are the MC data and the lines the fits obtained by means of
eq.~(\ref{Binder}).

\item \textbf{figure 4:} Variations of $\ln < \delta {\cal M}^2> $
 at $T^*_c(L)$ as a function of $\ln L$. The slope of the linear least square
 fit is $2\beta / \nu=1.04$.
 
\item \textbf{figure 5:} Variations of the total specific heat at constant 
volume
$C_V/V$ and the contribution $C_{\mu}/V$ with temperature along the locus 
$\chi_{NNN}=0$ at volumes  $V/\sigma^3=5000,10000,20000$, and $40000$
(from left to right).

\end{itemize}

%%%%%%%%%%%%%%%%%%%%%%%%%%%%%%%%%%%%%%%%%%%%%%%%%%%%%%%%%%%%%%%%%%%
\newpage
%\section{References}

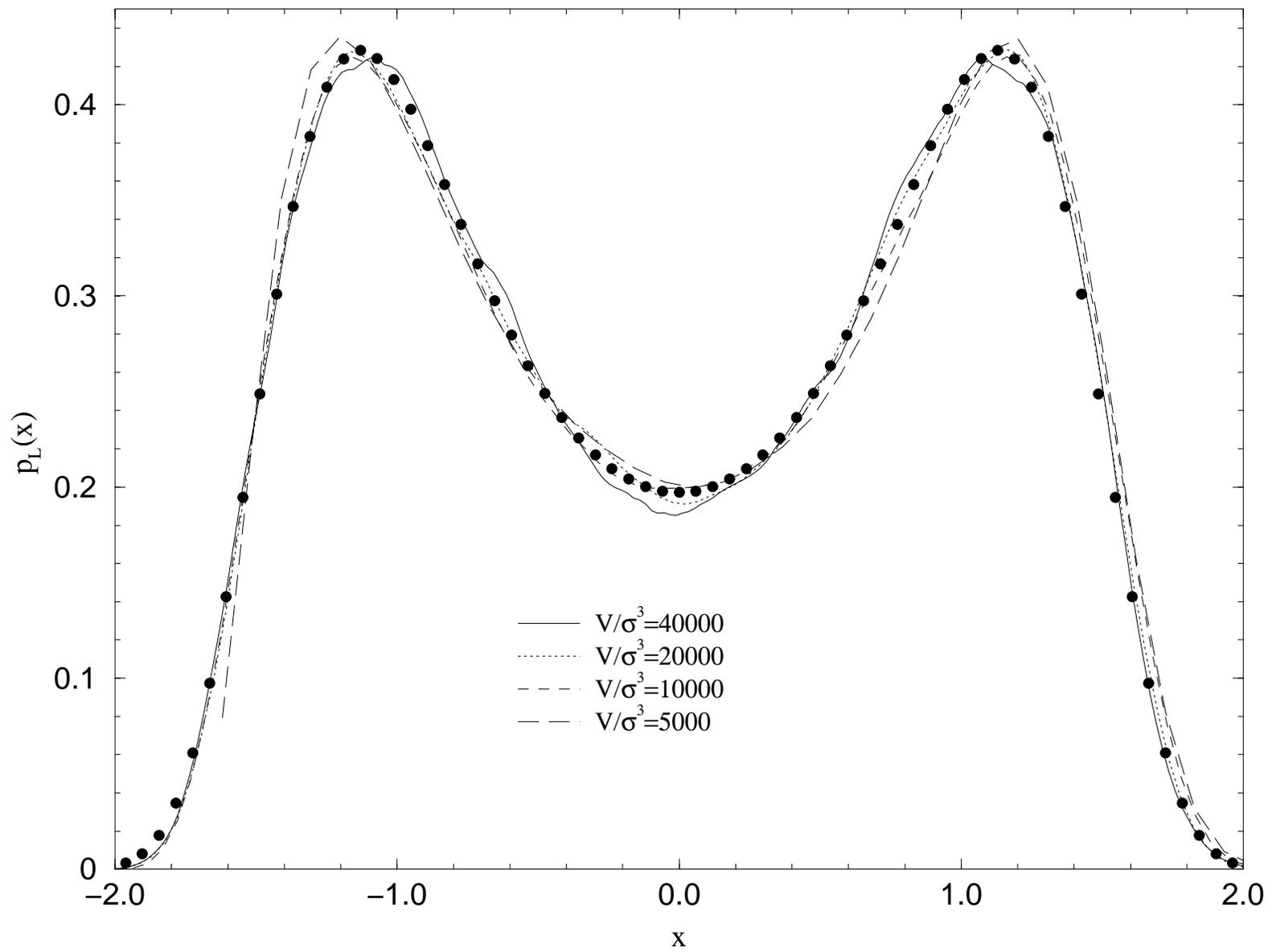
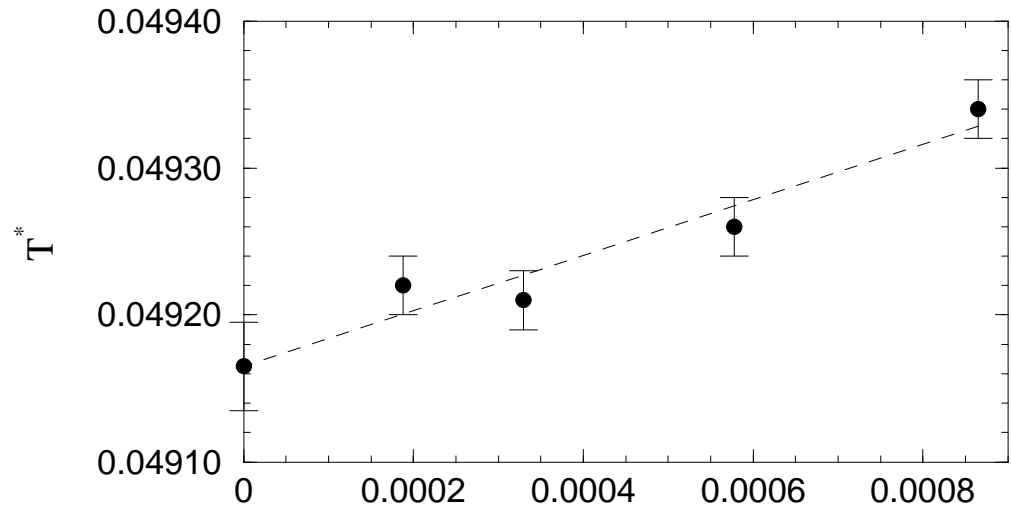
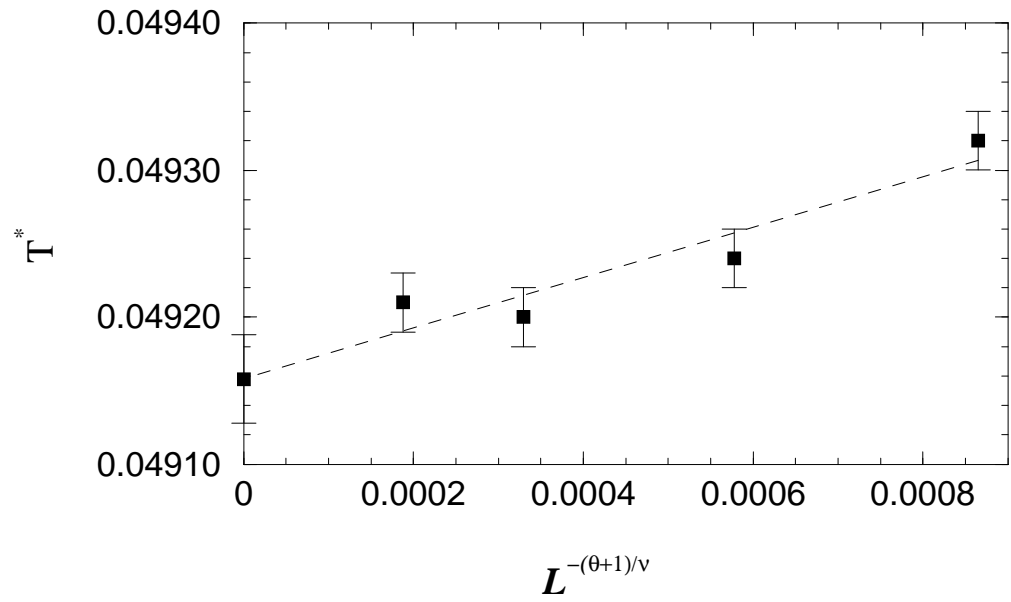
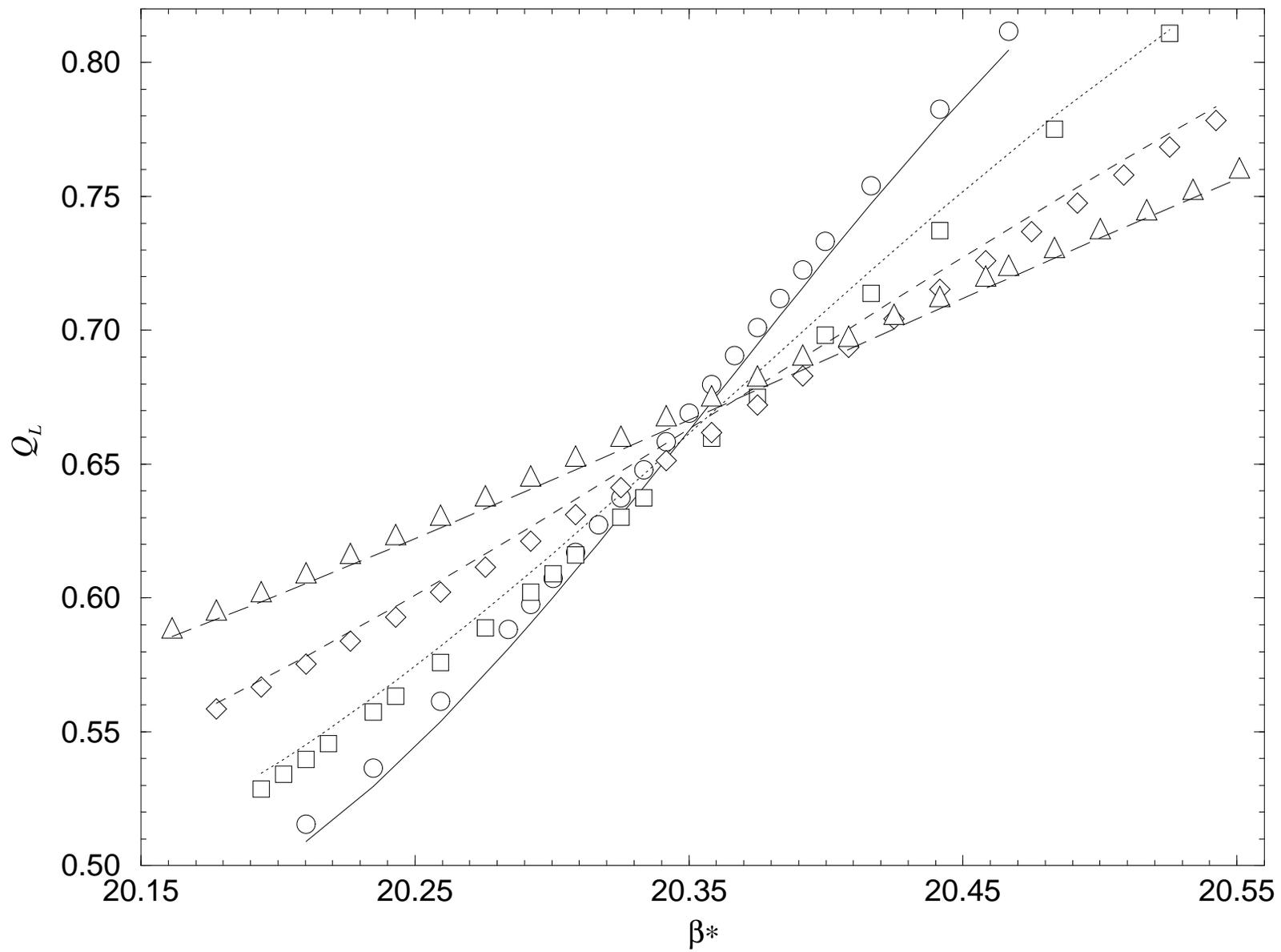
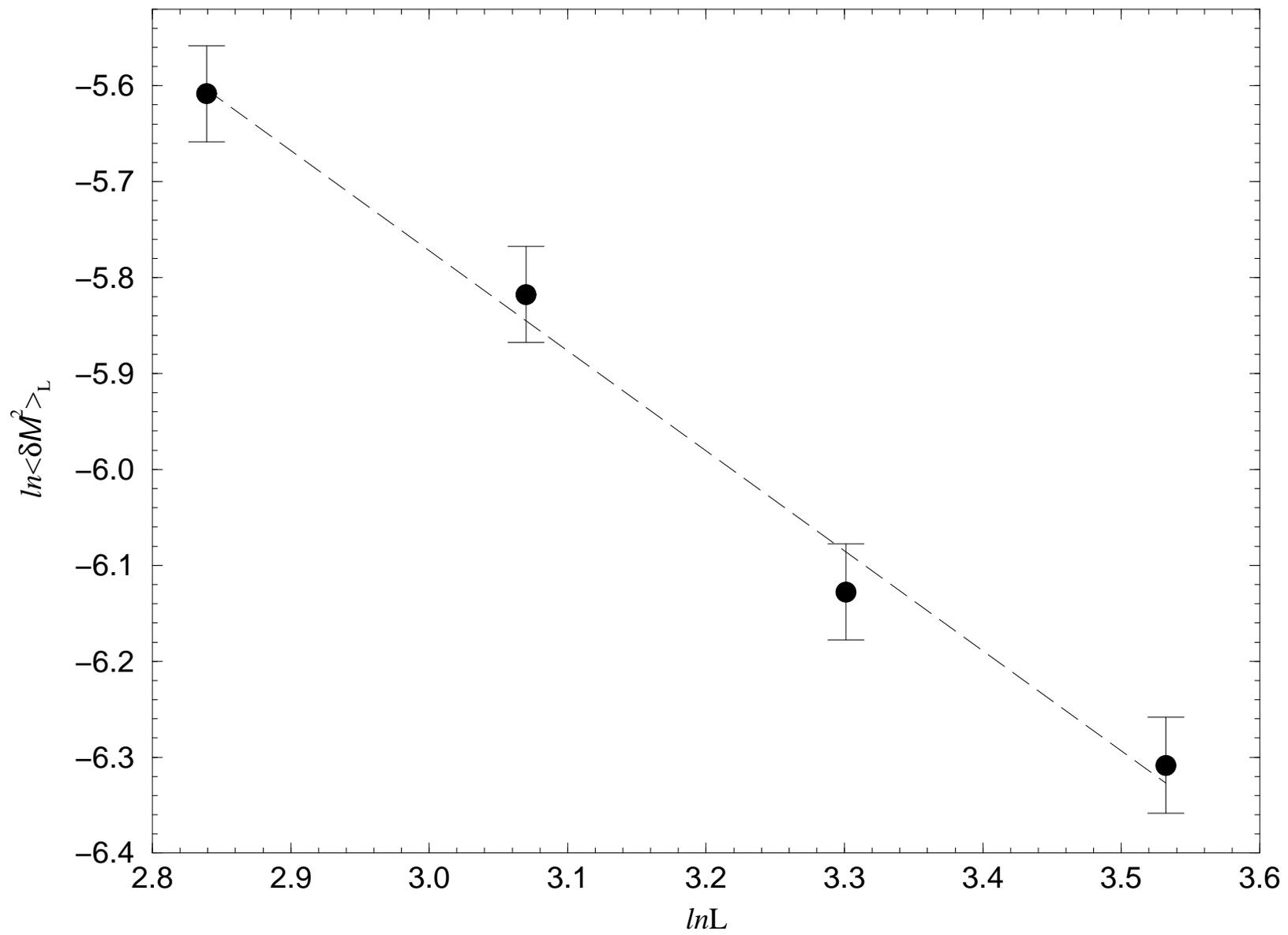
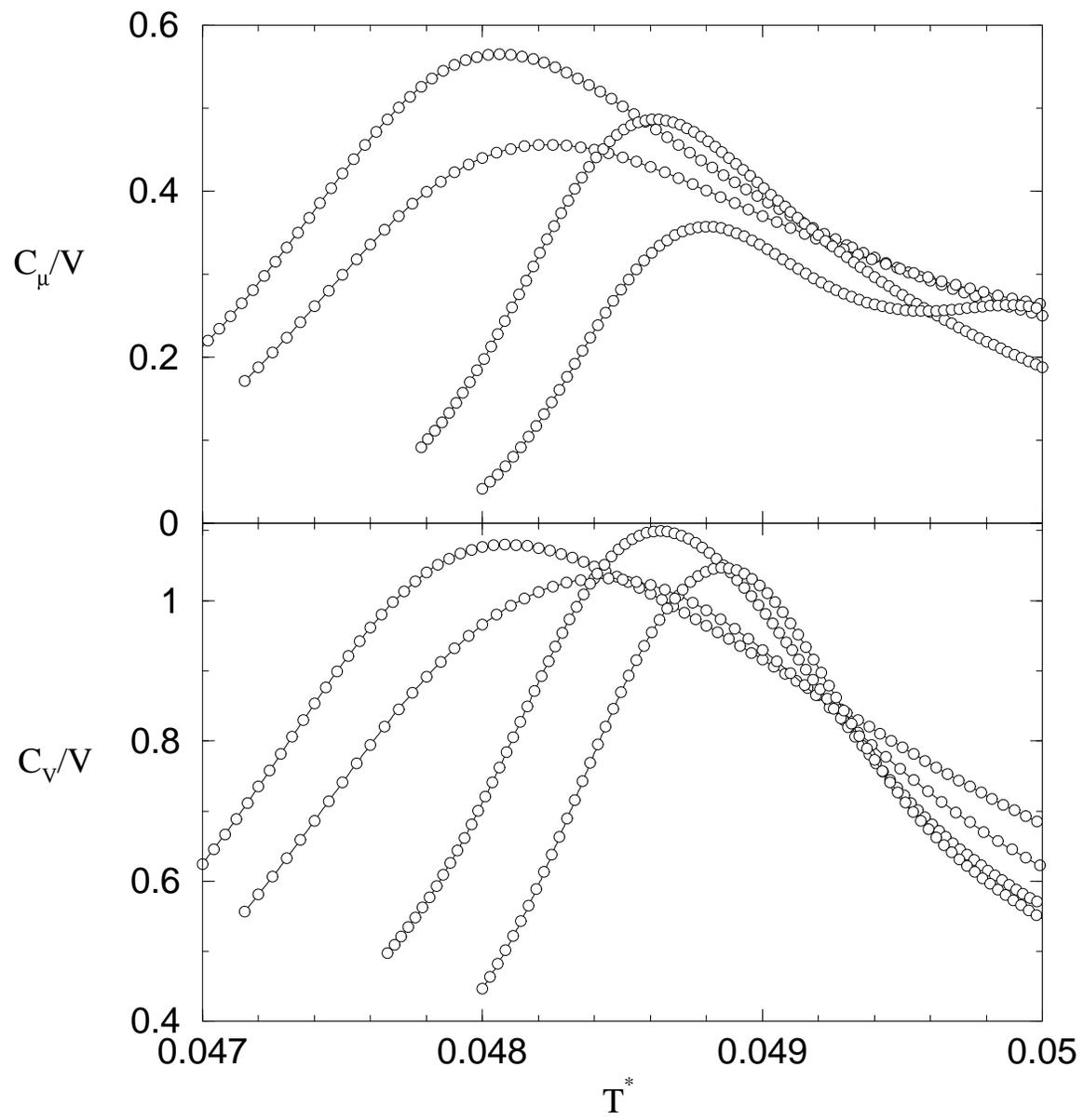
\end{document}